W.Pernice, C. Xiong, C. Schuck and H.Tang

# Second harmonic generation in phase matched aluminum nitride waveguides

W. H. P. Pernice[1*], C. Xiong[1], C. Schuck[1] and H. X. Tang[1†]

[1]*Department of Electrical Engineering, Yale University, New Haven, CT 06511, USA*

We demonstrate second order optical nonlinearity in aluminum nitride on insulator substrates. Using sputter-deposited aluminum nitride thin films we realize nanophotonic waveguides coupled to micro-ring resonators that simultaneously support cavity resonant modes for both visible and IR light. By using phase matched ring resonators, we achieve efficient second-harmonic generation and are able to generate up to 0.5μW of visible light on the chip with a conversion efficiency of -46dB. From the measured response we obtain a second order non-linear susceptibility ($\chi^2$) of 4.7pm/V. Our platform provides a viable route for realizing wideband linear and nonlinear optical devices on a chip.

---

[*] Current address: Institute of Nanotechnology, Karlsruhe Institute of Technology, 76133 Karlsruhe, Germany.

[†] Email: hong.tang@yale.edu

W.Pernice, C. Xiong, C. Schuck and H.Tang

Second order optical nonlinearity ($\chi^2$) in crystals such as lithium niobate (LiNbO$_3$) and potassium titanyl phosphate (KTP) has been widely exploited in modern optics for frequency doubling [1], parametric down conversion [2] and sum-/difference- frequency generation [3]. While optical nonlinearity is of importance for classical non-linear optics it is also an essential resource for the generation of non-classical light and applications in quantum optics [4, 5]. Recent development in photonics miniaturization has allowed the integration of large number of optical components on an integrated platform. In particular silicon photonics has emerged as the most promising platform for arranging optical devices in a scalable fashion [6,7]. However, the centrosymmetric lattice structure of unstrained silicon does not permit second-order optical nonlinearity. An approach to overcome this limitation is to bonding nonlinear crystalline materials with large intrinsic $\chi^2$ nonlinearity onto silicon substrate [8]. However, the bonding process required for substrate preparation can lead to film-uniformity issues. The obtainable quality factor in optical microresonators is further limited due to the remaining surface and interface roughness.

Here we present an integrated nonlinear optic system based on polycrystalline aluminum nitride (AlN) thin films on silicon substrates. AlN has been previously reported as a suitable material for SHG [9-11]. By relying on sputter-deposition, wafer-scale substrates with good uniformity can be fabricated for subsequent nano-patterning. Because AlN offers high second order nonlinearity [12], efficient wavelength conversion can be achieved on a chip. Since sputter-deposited AlN films have previously shown to provide good second-order nonlinear coefficients when deposited onto silicon substrates [13], we employ phase-matched ring resonators to demonstrate frequency up-conversion from 1550nm into the visible wavelength range.

W.Pernice, C. Xiong, C. Schuck and H.Tang

In order to allow for efficient wavelength conversion in second-order nonlinear materials, energy and momentum conservation must be fulfilled. In the case of nano-photonic waveguides, the phase-matching condition implies that the wave-vectors of the pumping optical mode and the second harmonic (SH) mode at twice the frequency must obey $k_{SH} = 2k_P$. This condition is equivalent to the requirement, that the effective indices of the pump $n_p$ and the SH mode $n_{SH}$ must be equal [14]. By engineering the dimensions of the nanophotonic waveguides this condition can be fulfilled by utilizing the geometric dispersion present in the waveguide [15].

We first employ finite-element simulations to obtain the dispersion relation of suitable waveguides and thus the effective indices. Results for various waveguide geometries are summarized in Fig.1. When the waveguide width is increased for a given waveguide height, the effective index of the fundamental mode in the telecoms window rises slower than the effective index of higher-order modes in the visible wavelength range. Therefore a cross-over between suitable modes can be found. In Fig.1(a) we show the cross-over points between the fundamental mode and the fifth and sixth order modes at 775nm. For a waveguide height of 330nm the phase-matched waveguide width is around 880nm. When thicker AlN layers are employed, the phase matched waveguide width is shifted towards wider waveguides. Here we chose a layer thickness of 330nm, which provides good coupling efficiency with our grating coupler approach and is thin enough for the etch selectivity of our reactive ion etching (RIE) process. In this configuration phase-matching is achieved between the fundamental mode at 1550nm and the fifth order mode at 775nm. The calculated modal profiles are shown in Fig.1(b). Sufficient spatial overlap between the guided modes ensures that energy transfer between the fundamental mode and the higher order modes can be obtained.

W.Pernice, C. Xiong, C. Schuck and H.Tang

We first investigate the phase-matching behavior of AlN nanostructures experimentally using straight waveguides. Highly c-axis oriented polycrystalline AlN thin films are sputter deposited onto silicon substrates with a buried oxide layer of 2.6μm thickness, thus yielding an AlN-on-Insulator (AOI) photonic substrate. Following the fabrication procedure described in [8], our devices are defined using electron-beam lithography on a Vistec EBPG 5000+ 100kV system. The patterns are transferred into the AlN thin films using inductively coupled reactive ion etching in $Cl_2$/$BCl_3$/Ar chemistry. Focusing grating couplers [16] are used to launch light into and out of the fabricated chip. In order to increase the interaction length of the pump light with the non-linear material the straight waveguides are arranged in a meander structure resulting in a total length of 1mm. An optical micrograph of a fabricated device is shown in Fig.2(a). To avoid loss occurring at the meander turns, we chose a bend radius of 50μm. The waveguide output is split into two ports: one leading to a grating coupler designed for a wavelength of 1550nm and the other leading to a second grating coupler designed for a wavelength of 775nm. The insertion loss of the grating couplers is approximately -13.5dB for 1550nm and -11.5dB for 775nm input light, respectively.

We couple pump light around 1550nm from a tunable diode laser into the device via the input grating coupler. In order to achieve efficient second-harmonic generation, the pump light is amplified using an erbium doped fiber amplifier (EDFA, Pritel FA-30). The output of the straight waveguide's 775nm coupler port is fed into a visible light spectrometer (Ocean Optics Jaz) for spectral analysis. Results are presented in Fig. (2b) for four different waveguide widths. We measure the detected SH count rate recorded with the spectrometer in dependence of wavelength for devices with different waveguide widths ranging from 845nm to 860nm. For each waveguide width a peak is observed at the corresponding wavelength where phase-matching is achieved.

The FWHM of the phase-matching peaks is on the order of 5nm in wavelength. We observe a linear increase in optimal phase-matching wavelength when the waveguide width is increased, with a slope of 0.6 as shown in Fig. 2(c). Thus by tuning the geometry of the waveguide we can tailor the second harmonic emission spectrum to a desired wavelength window.

The efficiency of the SH process can be significantly improved by employing optical cavities. Here we use AlN optical ring resonators with a radius of 130μm, coupled to two sets of waveguides, as shown in Fig.3(a): a 1000nm wide pump waveguide laid out for single-mode propagation at 1550nm (in the top half of the image) and a second 200nm wide collecting waveguide for operation around 775nm (in the bottom half of the image). Although it is straightforward to achieve critical coupling for IR light, it is much more difficult to realize critical coupling at shorter wavelengths due to weaker evanescent coupling between waveguides. In order to increase the coupling of the 775nm-collecting waveguide to the ring, the waveguide is wrapped around the ring using a pulley structure [17]. The coupling regime is further illustrated in the scanning electron microscope image in Fig.3(b). Using a narrow waveguide design for the 775nm output port also guarantees that the ring is not loaded at 1550nm wavelength and thus does not compromise the optical quality factor at the pumping wavelength. When measuring the transmission spectrum of the device at 1550nm input optical resonances with a free spectral range of 1 nm can be readily observed as illustrated in Fig.3(c). Critical coupling is achieved at a gap of 650nm where we observe extinction ratios of up to 30dB and optical quality factors on the order of 100,000. When considering devices with a wider coupling gap, the resonators are operated in a weakly coupled regime. In this case we find improved quality factors up to 200,00, which are expected to correspond more closely to the intrinsic quality factor of the device. From the fitted resonance curve [18] we extract propagation loss of 1.7dB/cm, which is on the same order as in silicon nanophotonic devices.

Having characterized the device spectral performance we measure the amount of generated second-harmonic light. As previously for the straight waveguide, pump light from a tunable diode laser (New Focus 6428) is passed through an EDFA to generate high optical input power at 1550nm wavelength. The amplified light is coupled into the device and the output from the 775nm ports is monitored with a low-noise photoreceiver (New Focus 2011). When the input wavelength is tuned to one of the optical resonances within the phase-matching window we observe strong SH light at the 775nm output port. For the ring we employ a waveguide width of 860nm in order to tune the phase-matching wavelength towards the low-loss propagation region of the feeding waveguide. When the pump wavelength is tuned all the way into the resonance we observe estimated power of 0.546µW at 775nm in the waveguide (or equivalently 40nW coupled out of the chip). The value is obtained by taking into account the loss due to the output grating coupler (-11.5dB), at an input optical pump power of 22mW on the waveguide. The propagating power in the feeding waveguide is corrected by the coupling loss at the input grating coupler (-13.5dB). From the measured results we obtain a peak conversion efficiency of -46dB (defined as the ratio of the micro-ring's SH output power relative to the input pump power), which is comparable to previously reported results in gallium nitride ring resonators [8]. The generated SH intensity translates into a normalized conversion efficiency, which we define as $\eta_n = P_{2\omega}/P_\omega^2 L^2 F^2$, where $L$ is the circumference and $F$ the optical finesse of the ring resonator. Because of the optical cavity the effective length of the device is enhanced by the finesse of the resonator. Using the free-spectral range (FSR) of the device of 1.004nm and the resonance linewidth of $\Delta\lambda = 140 pm$ to obtain the optical finesse as $F = FSR/\Delta\lambda$, we find a normalized

W.Pernice, C. Xiong, C. Schuck and H.Tang

conversion efficiency of -24.7dBW$^{-1}$cm$^{-2}$. While the efficiency is lower than reported values for periodically poled lithium niobate waveguides[19,20], the . When the pump light is tuned into one of the resonances, the SH light generated inside the resonator is strong enough to scatter off the residual surface roughness and its out-of-plane emission is visible on a CCD camera in top view. An optical micrograph of a glowing ring is shown in Fig.3(d).

When we measure the intensity of the SHG light in dependence of pump power on the waveguide as shown in Fig.3(e) a clear quadratic dependence can be observed. The best fit with a power law reveals an exponent of 1.94±0.07, which is in good agreement with the expected quadratic dependence on the input power. From the measured results we can extract the second order nonlinear susceptibility $\chi^{(2)}$ using the expression [21]

$$\chi^{(2)} = \sqrt{\eta \Gamma \frac{8 n_P^2 n_{sh} c^3 \varepsilon_0 A_P^2}{F_P^2 F_{sh} \omega_P^2 L^2 P_P A_{sh}}}$$

Here $\omega_p$ is the pump frequency, $\varepsilon_0$ is the permittivity of free space, $c$ is the speed of light in vacuum, $n_i$ is the effective refractive index for the pump and second-harmonic modes, $A_i$ is the mode area, $L$ is the ring circumference, $\eta$ denotes the conversion efficiency and $P_p$ is the pump power in the waveguide. $\Gamma$ is the modal overlap between the fundamental and second-harmonic modes and $F_i$ denotes the field enhancement factors on resonance. Using COMSOL Multiphysics to determine the modal profiles and the transmission spectra of the ring to determine the field enhancement factors (30 for the fundamental (pump) mode and 1.44 for the second-harmonic), we find a value of $\chi^{(2)}$ of 4.7±3pm/V, which is on the same order as previously reported results [13,22].

Due to its remarkably wide transparency window, spanning from the mid-IR all the way to visible/UV wavelengths (220 nm~13.6 µm), AlN is a promising candidate for non-linear optical applications. Since AlN thin films can be sputter-deposited onto suitable substrates, wafer-scale fabrication is feasible. As shown here high quality photonic components can be fabricated from AlN with standard nano-fabrication routines and are thus compatible with traditional integrated circuit design. The combination of AlN's high second order nonlinearity with its excellent optical properties provides a route towards a multitude of wideband optical applications based on our AOI platform. Low propagation loss makes it possible to generate large-area photonic circuits for applications in non-linear and traditional photonic devices.

This work was supported by DARPA/MTO. W.H.P. Pernice acknowledges support by the DFG grant PE 1832/1-1. C.X. acknowledges support the NSF grant through MRSEC DMR 1119826. H.X.T acknowledges support from a Packard Fellowship in Science and Engineering and a CAREER award from the National Science Foundation. We want to thank Dr. Michael Rooks and Michael Power for their assistance in device fabrication.

W.Pernice, C. Xiong, C. Schuck and H.Tang

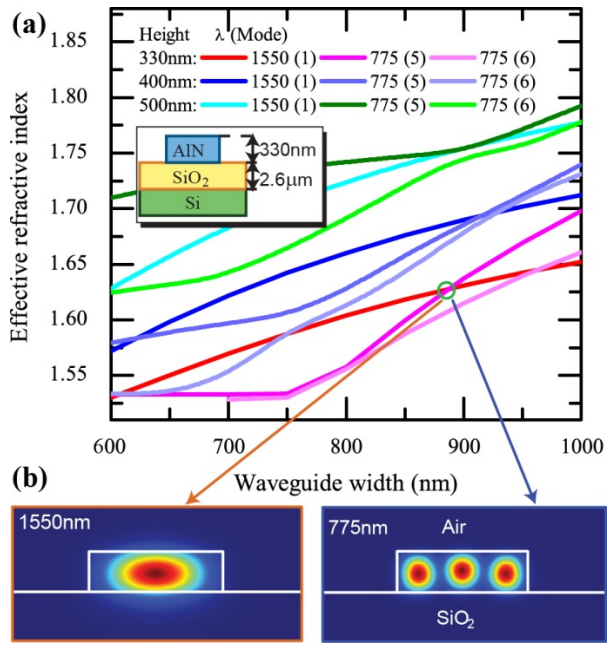

Fig. 1. (a) The calculated dispersion relations for optical modes at 775nm and 1550nm for various film thicknesses. Phase matching between higher order modes at 775nm and 1550nm can be achieved when the waveguide width is adjusted accordingly. In our case the phase-matching condition is fulfilled between the fifth-order mode at 775nm and the fundamental mode at 1550nm (the modal profiles are illustrated in (b)).

W.Pernice, C. Xiong, C. Schuck and H.Tang

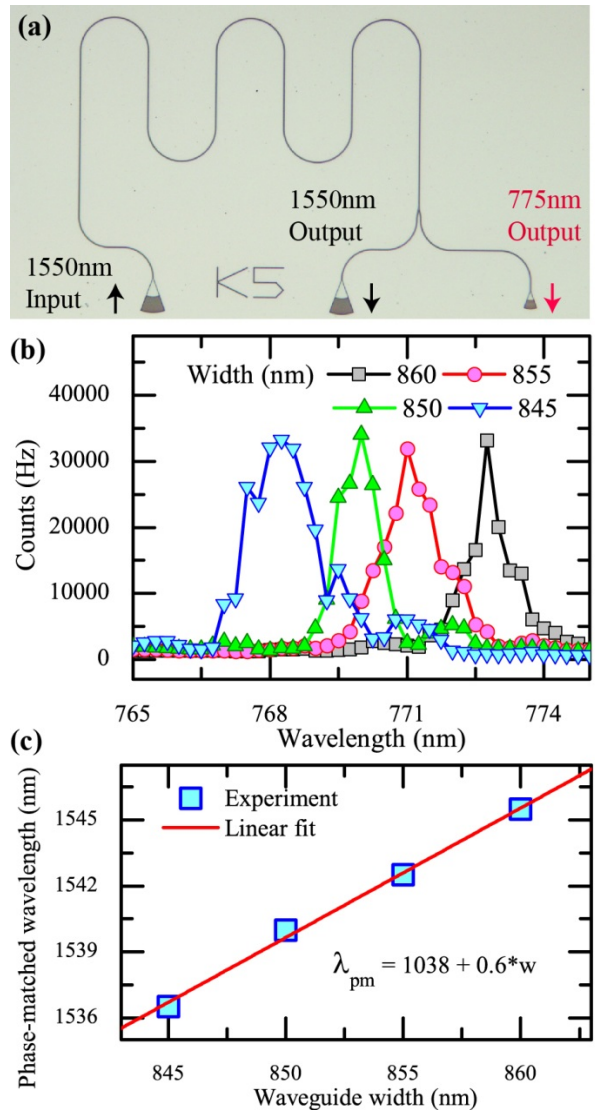

Fig.2. (a) An optical micrograph of a fabricated device for on-chip SHG. In order to increase the interaction length the waveguide is arranged in meander form. (b) SHG spectra obtained for four different straight waveguides with a width ranging from 845nm to 860nm. Phase-matching is achieved over a bandwidth of roughly 5nm. (c) The optimal phase-matched wavelength in dependence of waveguide width. The linear fit shows a slope of 0.6, allowing for adjusting the phase-matching wavelength via the waveguide width.

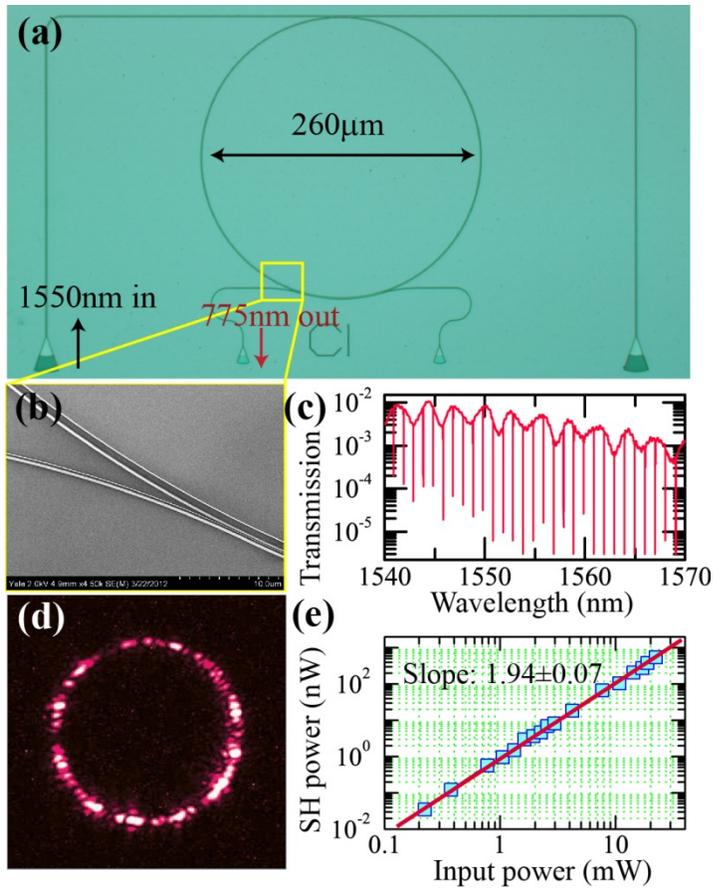

Fig.3. (a) An optical micrograph of a fabricated device for measuring the optical Q of AlN resonators. The ring resonator is coupled to two sets of waveguides for light at 775nm (bottom) and 1550nm (top). (b) A SEM image of the coupling region between the ring resonator and the 775nm outcoupling waveguide. (c) A microring glowing with SH generated light at 775nm. (d) The measured transmission spectrum of the device shown in (a) at near-infrared wavelengths. (e) The measured SHG power in dependence of pump power (both on the waveguide). A quadratic fit to the experimental data confirms the second-order non-linear relationship with a slope of 1.94±0.07.